\documentclass{article}

\usepackage{microtype}
\usepackage{graphicx}
\usepackage{subcaption}
\usepackage{booktabs}
\usepackage{array}
\usepackage{hyperref}

\usepackage[accepted]{icml2026}
\usepackage{amsmath}
\usepackage{amssymb}
\usepackage{mathtools}
\usepackage{amsthm}
\usepackage[capitalize,noabbrev]{cleveref}
\usepackage{placeins}

\newcommand{\dataset}{\mathcal{D}}

\newcommand{\package}{\mathcal{E}}
\newcommand{\auditor}{\mathcal{A}_{\theta}}

\icmltitlerunning{White Box Evidence Packages for Policy Audit Reports}

\begin{document}

\twocolumn[
  \icmltitle{White Box Evidence Packages for Policy Audit Reports}

  \begin{icmlauthorlist}
    \icmlauthor{Seunghyun Yoo}{grail}
  \end{icmlauthorlist}
  \icmlaffiliation{grail}{Governance and Responsible AI Lab}
  \icmlcorrespondingauthor{Seunghyun Yoo}{yoo236@purdue.edu}
  \icmlkeywords{mechanistic interpretability, AI auditing, technical AI governance, sparse autoencoders, white box evaluation}

  \vskip 0.3in
]

\printAffiliationsAndNotice{}

\begin{abstract}
As AI governance moves from benchmark scores toward auditable oversight, a central question is how reviewers can tell whether an LLM generated audit report is actually supported by evidence. This paper studies that question in passage anchored policy audits, where a report must interpret a given policy passage and cite evidence for its claims. We introduce a controlled evaluation framework that holds the passage, rubric, and auditor model fixed while changing only the evidence interface supplied to the auditor. Across 60 AGORA policy cases, we generate 600 structured reports under ten evidence conditions, including passage based evidence, internal model evidence, a hybrid package, and a shuffled control that preserves evidence format while breaking case relevance. We use five reviewer gold and diagnostic checks as a validation layer for correctness, passage grounding, diagnostic usefulness, and evidence misuse. The results show that internal evidence changes how reports cite and reason about evidence, but more internal citations do not by themselves make a report more valid. A separate behavior locked residual stream patching diagnostic finds a narrow, prompt sensitive candidate causal localization, while reports readily reuse broader readable labels and token directions. The hybrid interface is a promising design signal, while the shuffled control exposes a key governance risk: reports can sound substantively plausible while citing irrelevant internal evidence. This study reframes internal model access as an evidence design problem for audit workflows, rather than as a guarantee of transparency.
\end{abstract}

\section{Introduction}

AI governance increasingly needs audit reports that are not only persuasive, but checkable. Regulatory audits, AI safety institute evaluations, DSA style platform audits, and assurance standards all push oversight toward claims that can be traced to evidence rather than only to model outputs or benchmark scores \citep{iaasb2013isae3000,solarova2026checkbox,holznagel2025dsa,reuel2024technical}. Black box access remains essential because it captures what external users can observe, but it does not by itself answer a harder question: whether an audit report used the right evidence for the claim it makes \citep{casper2024blackbox}. This paper studies that evidence problem for LLM assisted policy audit reports.

We focus on passage anchored policy audits. In this setting, a report must interpret a given policy passage, cite supporting evidence, and make uncertainty visible to a later reviewer. The passage is the source of truth; the report is not a legal compliance decision, a human auditor experiment, or a capability audit where model internals may be the primary object of study. The key experimental question is simple: if the passage, rubric, and auditor model stay fixed, how does the evidence interface change the report?

The evidence interfaces compare surface evidence, internal model evidence, hybrid evidence, and a relevance broken control. Here, white box evidence means auditor facing entries produced from fixed interpretability tools applied to Gemma 2 2B. These entries are not treated as trusted explanations. They include text, source information, and caveats that the auditor may cite in a structured report. The report is therefore the unit of analysis: its findings, passage spans, evidence citations, confidence, and misuse of evidence.

\begin{figure*}[t]
    \centering
    \includegraphics[width=0.98\textwidth]{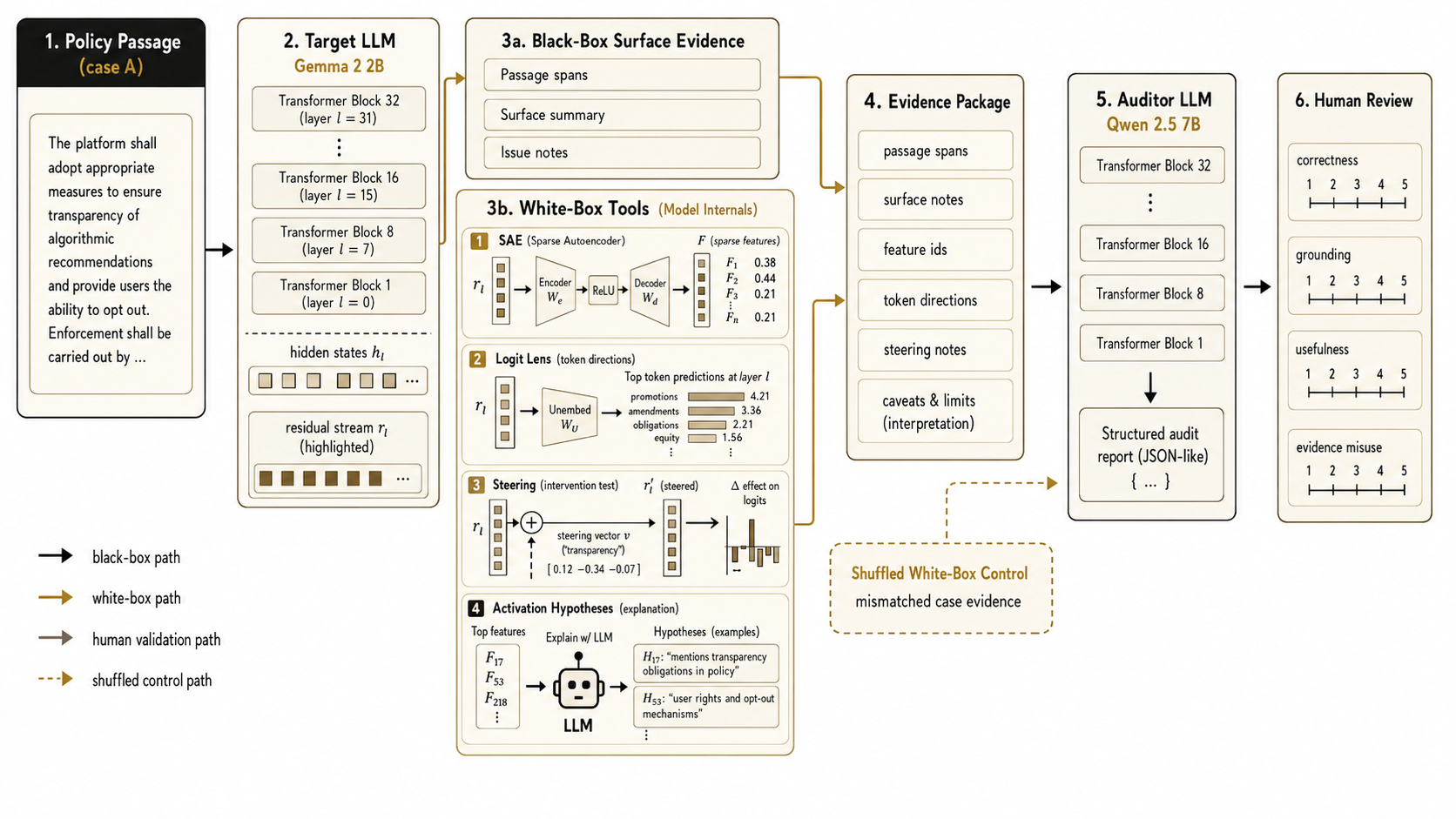}
    \caption{Evidence package audit design from policy passage to target model evidence, evidence package, auditor report, and human review. Internal states are summarized through SAE, logit lens, steering, and activation explanation surrogate records. The unit of comparison is the evidence interface, not a retrieval system, and the shuffled control preserves interface form while breaking case relevance. The displayed block sequence is schematic and indicates data flow rather than exact model depth.}
    \label{fig:package-design}
\end{figure*}

\Cref{fig:package-design} summarizes the design. We construct 60 AGORA policy cases, generate 600 structured reports with a local Qwen 2.5 7B auditor across ten evidence conditions, and use five reviewer gold and diagnostic checks as a validation layer. This validation review covers four interfaces over all 60 cases: a strong black box surface baseline, combined white box evidence, hybrid surface and white box evidence, and a shuffled white box relevance control that keeps the evidence format while pairing internal evidence with the wrong case.

The results support three claims. First, internal evidence changes citation behavior, but citation uptake is not evidence validity. Second, hybrid packages are a promising interface in this design, but their advantage is a design signal rather than a causal proof because the hybrid package is larger and more varied. Third, the failure mode is shallow use rather than non use: the auditor repeatedly cites readable SAE labels and token direction summaries, while a behavior locked residual patching diagnostic finds a narrow, prompt sensitive answer readout signal at the final prompt token. The contribution is therefore not a claim that white box access generally improves audits. It is a reproducible framework for evaluating internal model evidence as part of a reviewable audit workflow. To support reproducibility, we provide code, prompts, evidence packages, final reports, validation review files, and paper figures at \url{https://github.com/GRAIL-center/agora-mi}.

\section{Related Work}

Audit access work motivates the need for evidence interfaces, but access is not the same as usable evidence. Technical AI governance work argues that oversight institutions need tools for access, assessment, verification, and accountability rather than only post hoc descriptions of model outputs \citep{reuel2024technical,reuelbucknall2024open,ojewale2025accountability}. Related audit access work argues that black box interaction can miss information needed to test developer claims and system risks, motivating controlled access to weights, activations, development evidence, or verification channels \citep{casper2024blackbox,waiwitlikhit2024trustless,harack2025verification}. Our study accepts that motivation but evaluates the next step: whether internal information can be packaged, cited, and checked in an audit report.

Audit practice makes evidence relevance central. Assurance standards and regulatory audit regimes ask auditors to connect conclusions to sufficient and appropriate evidence, while early DSA audit critiques show that formal audit reports can remain weak when audit criteria, evidence relevance, and systemic risk tests are underspecified \citep{iaasb2013isae3000,solarova2026checkbox,holznagel2025dsa}. Access scholarship similarly warns that API and platform restrictions can create audit blind spots even when transparency mandates exist \citep{burnat2025accountability,cen2024transparency}. Stakeholder first transparency work argues that explanations should be tailored to the decision maker and use context rather than treated as generic disclosures \citep{bell2022stakeholders}. We apply these ideas to LLM generated audit reports: an evidence package should make evidence available, but it should also make relevance and misuse reviewable.

Mechanistic interpretability provides candidate sources of internal evidence, but readable tool outputs still need downstream validation. Sparse autoencoders can turn residual stream activations into feature hypotheses \citep{cunningham2024sae,bricken2023monosemanticity,templeton2024scaling,lieberum2024gemmascope}; logit lens methods can summarize intermediate token directions \citep{belrose2023tunedlens}; steering and sensitivity methods can test how internal directions affect model behavior \citep{cyberey2026steering}; and automated explanation methods can convert activations into natural language hypotheses \citep{bills2023neurons,activationoracles2025}. Recent evaluation work warns that readable labels and tool outputs should be tested through downstream use, ablations, and controls rather than treated as explanations by default \citep{karvonen2025saebench,sheshadri2026auditbench}. AuditBench is closest in spirit because it evaluates multiple white box tools as audit instruments; our setting differs by evaluating the downstream policy report artifact while holding the same passage and rubric fixed.

The shuffled control draws on known LLM context reliability failures. Prior work shows that language models can be distracted by irrelevant context, fail to use long context reliably, over rely on retrieved passages, and exhibit sycophantic agreement patterns under social or evidential pressure \citep{shi2023distracted,liu2024lost,yoran2024robustrag,sharma2023sycophancy}. We do not claim to newly discover that LLMs can overuse irrelevant evidence. Our contribution is to instantiate that risk in governance facing white box evidence packages and measure whether reports rely on internal evidence that belongs to the wrong case.

\section{Audit Task and Evidence Interfaces}

Each case asks the auditor to produce a reviewable report from a policy passage. The input contains a policy passage, a fixed audit rubric, and one evidence package. The hidden gold brief is withheld from the auditor and used only for validation review. The output report must identify policy relevant issues, cite supplied evidence entries and exact passage spans, record uncertainty, and avoid claims not supported by the package.

This task fixes the authority structure of the audit. The passage is the object being interpreted, and white box evidence is auxiliary evidence about how a target model represents or reacts to that passage. This differs from a capability or behavioral audit where there may be no passage level ground truth and internal evidence may be the main object of investigation. Our results should therefore be read as evidence about LLM assisted passage grounded policy audit reports, not as a ranking of white box tools for all audit settings.

The experimental control is deliberately simple: the passage, rubric, and auditor stay fixed, and only the evidence package changes. Every package contains stable evidence entries, source information, text, and caveats. The auditor is asked to cite an entry only when it directly supports a finding. This design separates three objects that audit reports often blur: the policy passage, surface observations about the passage, and internal evidence about the target model.

Formally, the report for case $i$ under condition $c$ is $r_{i,c}=f(x_i,\mathcal{R},\auditor,\package_{i,c})$, where $x_i$ is the passage, $\mathcal{R}$ is the fixed rubric, and $\package_{i,c}$ is the condition specific evidence interface. We hold $x_i$, $\mathcal{R}$, and $\auditor$ fixed when comparing conditions and vary only $\package_{i,c}$.

\Cref{tab:conditions} lists the ten evidence conditions. Every interface includes the policy passage and fixed rubric; the condition names describe the additional evidence shown to the auditor. The main narrative focuses on four primary interfaces. Black box surface evidence is the passage anchored baseline built from selected spans, issue notes, and confidence annotations. Combined white box evidence collects all four internal evidence views. The hybrid interface appends the combined white box package to the surface anchors. The shuffled white box relevance control preserves the combined package format but draws its internal evidence from another case, testing whether plausible looking evidence is treated as authoritative when it should not be.

\begin{table}[h!]
    \centering
    \caption{Evidence conditions.}
    \label{tab:conditions}
    \scriptsize
    \setlength{\tabcolsep}{3pt}
    \begin{tabular}{@{}p{0.36\columnwidth}p{0.56\columnwidth}@{}}
        \toprule
        Interface & Auditor facing contents \\
        \midrule
        Passage only baseline & Passage and rubric only. \\
        Black box surface evidence & Surface summary, selected spans, issue notes, and confidence. \\
        Sparse autoencoder evidence & Feature activations, layers, labels, spans, and caveats. \\
        Logit lens evidence & Intermediate token direction summaries. \\
        Steering sensitivity evidence & Directional sensitivity and output or score deltas. \\
        Activation explanation surrogate & Natural language activation hypotheses. \\
        Combined white box evidence & Sparse autoencoder, logit lens, steering, and activation explanation evidence together. \\
        Hybrid surface and white box evidence & Black box surface evidence plus combined white box evidence. \\
        Raw AutoInterp sparse autoencoder evidence & Raw AutoInterp labels and feature references. \\
        Shuffled white box relevance control & Combined white box evidence drawn from another case. \\
        \bottomrule
    \end{tabular}
\end{table}

The remaining conditions are diagnostic single tool interfaces. They expose SAE features, logit lens token directions, steering sensitivity records, activation explanation surrogates, or raw AutoInterp labels in isolation. These conditions help interpret tool behavior, while the four primary interfaces receive the full validation review used to check report correctness, grounding, usefulness, and misuse. Tool construction details and formulas are provided in the appendix.

\FloatBarrier

\section{Experimental Setup}

\paragraph{Cases and models.}
We use 60 policy cases drawn from AGORA, a corpus of AI governance and regulatory material \citep{arnold2024agora}. The cases come from 11 source documents and include obligations, rights, exceptions, enforcement mechanisms, and governance procedures. The interpreted target model is Gemma 2 2B, and the auditor is a local Qwen 2.5 7B model run with structured outputs. For each condition, the auditor receives only the passage, rubric, and assigned evidence package. It does not see the hidden gold brief and is not told that the shuffled white box relevance control contains evidence from another case.

\paragraph{Evidence generation.}
The package generator is deterministic given the selected case set. SAE items use Gemma Scope residual stream SAEs and human revised AutoInterp labels where available. Logit lens items use Gemma 2 2B layers 6, 12, 18, and 24 at audit cue token positions selected from the passage. Steering items use layers 12, 18, and 24 with a fixed obligation and risk contrast prompt set. The activation explanation condition is a conservative surrogate derived from selected SAE evidence and label information; it tests the interface role of natural language activation hypotheses, not a validated vector decoder.

\paragraph{Residual stream patching diagnostic.}
Because evidence uptake does not establish causal localization, we run a separate behavior locked residual stream patching diagnostic on Gemma 2 2B. This post hoc diagnostic is not an evidence item supplied to the auditor. It patches clean residual states into corrupted forced choice microtasks across deontic force and authority assignment, measures recovery of the correct answer margin, and compares candidate layer and token regions with random regions. The main layer plot uses the direct prompt shell and canonical answer order; prompt shell and answer order controls test whether the localization is stable. Full microtask construction, recovery, gate, and control details are provided in \cref{sec:appendix-patching}.

\paragraph{Report normalization.}
Every response is normalized into a structured report with findings, passage spans, cited evidence, internal evidence use, confidence, and cautionary notes. Structural checks verify that each report can be read, contains findings, records confidence consistently, and cites only evidence entries that were actually supplied. These checks establish that the pipeline produced inspectable reports, not that the reports are accurate.

\paragraph{Validation review.}
This review is used as a validation layer, not as the object of the study. Five reviewers contribute across gold writing, verification, diagnostic scoring, and agreement checks. First, reviewers write blind gold briefs from the policy passage and source information before seeing model reports or evidence packages. Second, reviewers score selected reports against those briefs to check whether cited evidence actually supports the passage grounded report. The validation review covers all 60 cases under the black box surface, combined white box, hybrid, and shuffled relevance interfaces, yielding 240 scored reports for correctness, span grounding, diagnostic usefulness, and evidence misuse. Agreement diagnostics use the original 20 case subset, and optional single tool diagnostic review covers sparse autoencoder, logit lens, steering sensitivity, activation explanation surrogate, and raw AutoInterp sparse autoencoder evidence on that subset. Lower is better only for misuse; the appendix gives the full review protocol.

Validation scores are summarized as condition means over the reviewed case set. Pairwise statements compare each condition with the black box surface baseline on the same cases and report better, tied, and worse counts; for misuse, lower scores are better. We use exact sign tests and Wilcoxon signed rank approximations as descriptive paired checks because the scores are ordinal and tied. Citation uptake is the mean number of evidence citations per report, and the shuffled warning counts relevance control reports with both high correctness and high evidence misuse. Full notation and score distributions are in the appendix.

\begin{table}[H]
    \centering
    \caption{Validation review axes. Scores use a 1 to 5 scale. Higher is better except for misuse.}
    \label{tab:review-axes}
    \scriptsize
    \setlength{\tabcolsep}{3pt}
    \begin{tabular}{@{}p{0.25\columnwidth}p{0.66\columnwidth}@{}}
        \toprule
        Axis & Review question \\
        \midrule
        Correctness & Does the report match the gold issue, actor, and obligation structure? \\
        Span grounding & Are major claims tied to concrete passage text? \\
        Usefulness & Would the report help an auditor notice the issue or gap? \\
        Evidence misuse & Does the report overread, overcite, or rely on irrelevant internal evidence? \\
        \bottomrule
    \end{tabular}
\end{table}

\FloatBarrier

\section{Results}

The results are organized around the evidence interface, not around the review process. We first test whether internal evidence changes uptake, then use validation review to check whether uptake corresponds to grounded support, then examine the hybrid design signal, the white box evidence the auditor actually cites, and the shuffled relevance failure.

\paragraph{RQ1: Does internal evidence change evidence uptake?}
Yes. The Qwen 2.5 7B auditor produced reports for all 600 condition and case combinations, and all reports were successfully normalized after retrying long responses. Structurally, the combined white box interface changes the report substantially: its reports cite 10.13 evidence entries per report versus 3.87 under black box surface evidence. However, the shuffled white box relevance control also yields 9.65 citations per report. Citation volume is therefore evidence uptake, not evidence validity.

\begin{figure*}[t]
    \centering
    \includegraphics[width=0.98\textwidth]{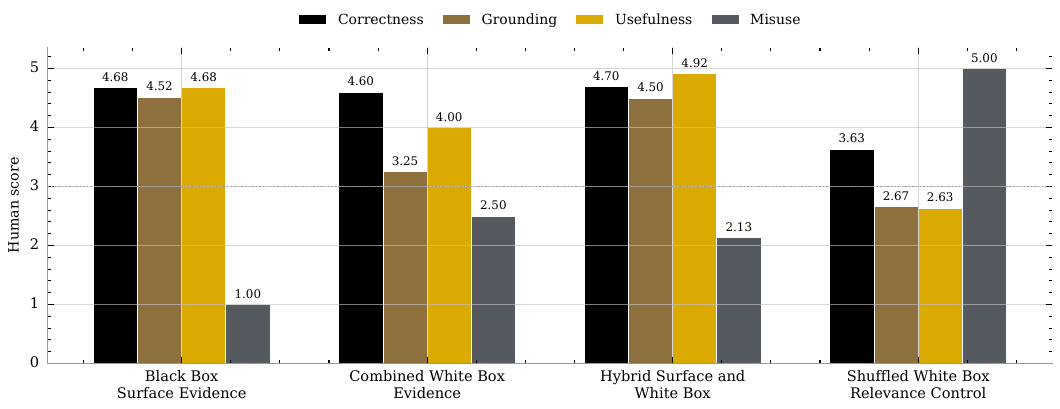}
    \caption{Validation review scores for black box surface evidence, combined white box evidence, hybrid surface and white box evidence, and the shuffled white box relevance control over the 60 case evaluation set. Scores use a 1 to 5 scale; higher is better except evidence misuse.}
    \label{fig:human-review}
\end{figure*}

\paragraph{RQ2: Does citation uptake correspond to grounded support?}
Not by itself. The validation review, conducted with five reviewers across gold brief and diagnostic stages, covers all 60 paired cases under black box surface, combined white box, hybrid, and shuffled relevance interfaces. \Cref{tab:human-results} and \cref{fig:human-review} summarize this check. Combined white box evidence preserves high average correctness (4.60 vs. 4.68 for the surface baseline), but it substantially weakens grounding (3.25 vs. 4.52), lowers usefulness (4.00 vs. 4.68), and increases misuse (2.50 vs. 1.00). The paired pattern is consistent with the averages: combined white box evidence is worse than the surface baseline on grounding in 53 of 60 cases and worse on usefulness in 41 of 60 cases, while evidence misuse is higher in all 60 cases.

\paragraph{RQ3: What does the hybrid interface suggest?}
The hybrid surface and white box interface is the strongest design signal in the validation review. It is nearly tied with the surface baseline on correctness (4.70 vs. 4.68) and grounding (4.50 vs. 4.52), while improving usefulness from 4.68 to 4.92. The paired usefulness effect is positive in 17 cases, tied in 40, and negative in 3. However, the hybrid interface also increases evidence misuse in all 60 cases, from 1.00 to 2.13 on average. This pattern is consistent with the hypothesis that hybrid evidence can work when surface passage cues anchor the report and internal evidence is used as a secondary signal. It is not a causal isolation of anchoring, because the hybrid package contains 22.97 evidence items on average, compared with 3.27 for black box surface evidence and 19.70 for combined white box evidence.

\paragraph{RQ4: Which white box evidence does the auditor actually cite?}
White box evidence is read shallowly rather than ignored. In raw report citations, the combined white box interface cites logit lens evidence most often, with 232 logit lens citations versus 68 sparse autoencoder citations. After normalizing by how many items are available, however, sparse autoencoder labels dominate: the combined interface cites 93.2\% of available sparse autoencoder entries, compared with 33.9\% of logit lens entries, 3.1\% of steering entries, and 2.8\% of activation explanation surrogate entries. The same pattern survives the shuffled relevance control, which cites 95.9\% of available sparse autoencoder entries despite the evidence coming from another case. The optional 20 case single tool review reinforces this warning: logit lens evidence and raw AutoInterp sparse autoencoder evidence each have mean grounding 1.25 and evidence misuse 4.00 (\cref{tab:appendix-single-tool}).

\begin{figure*}[t]
    \centering
    \includegraphics[width=0.98\textwidth]{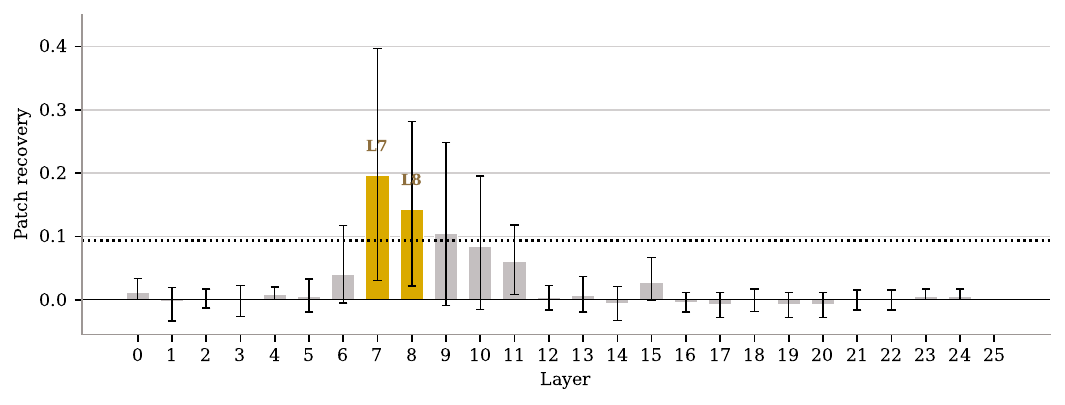}
    \caption{Residual stream activation patching diagnostic for deontic force under the direct prompt shell and canonical answer order. Bars show mean patch recovery by layer at the final prompt token; error bars show 95\% bootstrap intervals, and the dotted line marks the empirical causal gate.}
    \label{fig:whitebox-patching-layer}
\end{figure*}

\paragraph{Residual stream patching.}
A separate residual activation patching diagnostic gives the stricter white box view \citep{vig2020causal,zhang2024activationpatching}. In the direct shell, canonical order run, deontic force passes the causal gate at the final prompt token at layer 7 (mean recovery $0.1968$, 95\% bootstrap interval $[0.0302,0.3968]$) and layer 8 ($0.1429$, $[0.0220,0.2819]$), against an empirical threshold of $0.0935$. Authority assignment has no passing region. The localization is rendering sensitive: the direct shell, swapped order control also passes at layer 10, while the analyst shell, canonical order control has no passing region. The mechanism level reading is therefore narrower than the report citations and is best interpreted as a prompt sensitive late decision or answer readout signal, not broad causal support. \Cref{fig:whitebox-patching-layer} shows the main direct shell, canonical order result, and \cref{tab:appendix-patching-controls} reports the controls.

\begin{figure*}[t]
    \centering
    \includegraphics[width=0.98\textwidth]{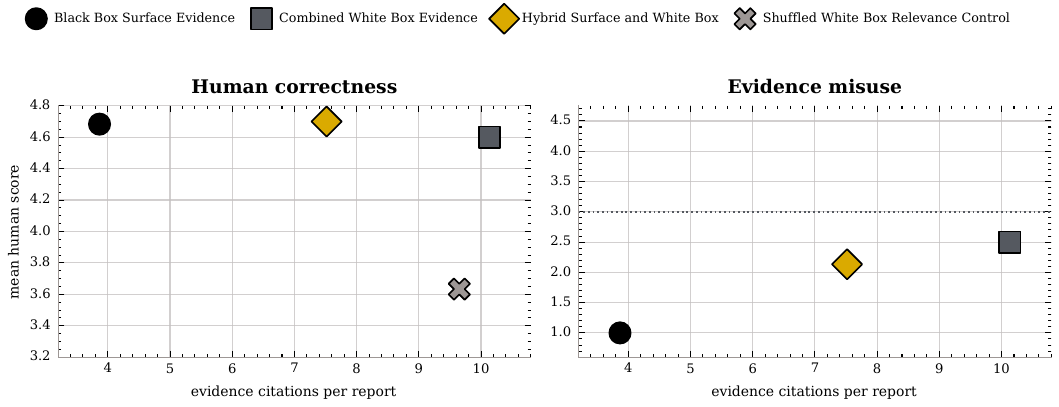}
    \caption{Citation volume plotted against validation correctness and evidence misuse. The x axis reports mean evidence citations per report over the 60 case run; the y axes report validation review scores over the same cases.}
    \label{fig:uptake-validity}
\end{figure*}

\paragraph{RQ5: Why is a shuffled control necessary?}
The shuffled white box relevance control is the strongest warning. It has lower average correctness than the surface baseline and hybrid interface, but it often remains substantively plausible because the auditor can still read the passage. Thirty nine of sixty shuffled control reports have correctness at least 4 while also receiving evidence misuse at least 4. Thus a report can be mostly right about the policy passage while invalidly citing internal evidence from another case. The auditor's own cautionary notes rarely catch this problem: only 3 of these 39 reports warn about weak evidence, and only one explicitly names an internal tool. \Cref{fig:uptake-validity} shows the same point from another angle: citation uptake cannot be used as a quality measure.

\begin{table}[H]
    \centering
    \caption{Validation review summary over 60 paired cases. Scores are means on a 1 to 5 scale. Higher is better except evidence misuse.}
    \label{tab:human-results}
    \small
    \begin{tabular}{@{}p{0.32\columnwidth}rrrr@{}}
        \toprule
        Condition & Correct & Ground & Useful & Misuse \\
        \midrule
        Black box surface evidence & 4.68 & 4.52 & 4.68 & 1.00 \\
        Combined white box evidence & 4.60 & 3.25 & 4.00 & 2.50 \\
        Hybrid surface and white box evidence & 4.70 & 4.50 & 4.92 & 2.13 \\
        Shuffled white box relevance control & 3.63 & 2.67 & 2.63 & 5.00 \\
        \bottomrule
    \end{tabular}
\end{table}

\begin{table}[H]
    \centering
    \caption{Paired deltas relative to black box surface evidence over 60 cases. The B/T/W column reports cases where the listed condition is better than, tied with, or worse than the surface baseline on the axis; for misuse, better means lower misuse, so positive deltas are worse.}
    \label{tab:paired-tests}
    \small
    \setlength{\tabcolsep}{4pt}
    \begin{tabular}{@{}lrrr@{}}
        \toprule
        Axis & Mean $\Delta$ & B/T/W & Sign $p$ \\
        \midrule
        \multicolumn{4}{@{}l}{\textit{Combined white box evidence}} \\
        Correct & $-0.08$ & 8/39/13 & .383 \\
        Ground & $-1.27$ & 0/7/53 & $<.001$ \\
        Useful & $-0.68$ & 0/19/41 & $<.001$ \\
        Misuse & $+1.50$ & 0/0/60 & $<.001$ \\
        \addlinespace[2pt]
        \multicolumn{4}{@{}l}{\textit{Hybrid surface and white box evidence}} \\
        Correct & $+0.02$ & 9/43/8 & 1.000 \\
        Useful & $+0.23$ & 17/40/3 & .003 \\
        Misuse & $+1.13$ & 0/0/60 & $<.001$ \\
        \addlinespace[2pt]
        \multicolumn{4}{@{}l}{\textit{Shuffled white box relevance control}} \\
        Correct & $-1.05$ & 0/6/54 & $<.001$ \\
        Misuse & $+4.00$ & 0/0/60 & $<.001$ \\
        \bottomrule
    \end{tabular}
\end{table}

\section{Qualitative Validation Evidence}

The aggregate result is easiest to interpret through actual policy passages. We use the case studies to check whether the numeric pattern means what the paper claims: hybrid evidence can help when passage anchors remain primary, while white box evidence can still mislead when internal labels are treated as substantive policy evidence.

The first case explains why the hybrid interface is useful. Articles 29 and 30 of the Chinese Algorithmic Recommendation Provisions contain two separate issues: confidentiality for supervision bodies and personnel, and a complaint or report mechanism for organizations and individuals. The black box surface baseline is correct and useful but not fully comprehensive on span grounding (5, 4, 5, 1). Combined white box evidence finds the related confidentiality theme but weakens grounding and usefulness relative to the passage anchored baseline (5, 3, 4, 2). The hybrid interface keeps surface evidence visible, so the report covers both confidentiality and complaint handling while using internal evidence secondarily (5, 5, 5, 2). The shuffled relevance control is partly correct but cites shuffled transparency evidence (4, 2, 3, 5). This is consistent with the interface claim: hybrid packages can keep white box evidence tied to the passage, but this case does not isolate anchoring from the larger evidence package.

The second case shows the limitation. The EU AI Act passage is a classification exception: an Annex III system is not considered high risk if it does not pose a significant risk under listed conditions, but profiling of natural persons remains always high risk. The black box surface baseline captures the not high risk exception and the profiling carve out (4, 4, 4, 1). Combined white box evidence preserves the broad classification topic but can still blur the legal category and cite weak internal evidence (4, 3, 4, 2). The hybrid interface improves anchoring but can still use imprecise low risk language (4, 4, 4, 2). The shuffled relevance control uses shuffled deployment evidence and misses part of the profiling override (3, 2, 2, 5). This is why the paper does not claim that hybrid evidence solves legal precision. It performs best on average, but legal category wording remains fragile.

Across the 39 shuffled relevance control reports with high correctness and high misuse, we observe three recurring patterns. First, the auditor can read the passage correctly while citing internal evidence from another case. Second, broad SAE or logit lens labels can look relevant to the policy family while failing to support the specific actor, obligation, or exception. Third, shuffled evidence can transfer the report toward plausible transparency, deployment, or risk themes that are not the relevant legal structure. These examples justify the shuffled control as more than a sanity check. A policy audit report can be useful at the surface level and still misuse the internal record. For governance uses, that is precisely the dangerous case: a reviewer who only checks whether the conclusion sounds plausible could miss that the cited internal evidence is irrelevant.

\FloatBarrier

\section{Discussion and Limitations}

The results support design lessons rather than a simple performance ranking. First, passage anchors should remain visible. The validation review repeatedly identifies broad SAE labels, generic logit lens directions, and non specific steering evidence as failure modes when they are detached from the concrete actor, obligation, exception, or remedy in the passage. Second, citation volume should not be used as a quality measure. Combined white box evidence and the shuffled relevance control both induce heavy citation behavior, but the validation review separates useful support from invalid support. Third, relevance controls are mandatory. If an LLM auditor can produce a plausible report under shuffled internal evidence, the interface risks transparency theater rather than reliable audit evidence.

The mechanistic interpretation is deliberately narrow. The auditor does not ignore the white box package: it cites SAE labels at very high per item rates and repeatedly reuses broad policy labels such as transparency, accountability, discrimination protection, and expert consultation. But the residual patching diagnostic shows why those citations should not be read as broad mechanistic support. In the direct shell, canonical order diagnostic, the localized deontic force signal appears at layers 7 and 8 at the final prompt token; the swapped order control also admits layer 10, and the analyst shell control has no passing region. This rendering sensitivity is more consistent with a late decision or answer readout state than a stable policy mechanism. The report interface exposes many more readable labels and token directions, so the overall pattern is a failure of evidence interpretation rather than a failure of evidence availability.

The intended governance use is narrow. The paper does not propose an automated legal compliance system, and it does not test whether human auditors would be fooled by shuffled evidence. Instead, it evaluates evidence interfaces for LLM assisted audit workflows that are reviewed by humans or institutions: reviewers need to know which evidence was available, what was cited, and whether irrelevant evidence would have produced similar confidence. For technical AI governance, the contribution is an operationalization of access and verification. For mechanistic interpretability, the contribution is an application level evaluation of tool outputs through downstream report use and negative controls.

Several limitations point to the next experiments. The practical hypothesis is that anchoring matters, but the current design does not isolate anchoring from evidence quantity and variety. In the validation review records, package size is positively associated with evidence misuse (Spearman $\rho=0.455$) but not with diagnostic usefulness ($\rho=-0.029$); this is not causal evidence, but it reinforces the need for volume controlled tests. A capped hybrid condition would separate anchoring from package size. A hybrid plus shuffled condition would test whether surface anchors protect against irrelevant internal evidence. Numeric only or label masked internal evidence would test whether readable labels drive over trust. A visible but not citable condition would separate evidence influence from citation behavior. Finally, running another auditor model or stricter prompt would test whether the pattern is Qwen specific.

The validation review uses five reviewers and covers all 60 cases for the four primary conditions, but it should still be read as a diagnostic study rather than a final benchmark. Agreement diagnostics are measured on the original 20 case subset rather than all 60 cases, and the sparse autoencoder, logit lens, steering sensitivity, activation explanation surrogate, and raw AutoInterp sparse autoencoder interfaces are human scored only on that subset. The target model is small, and the auditor is one local model. The activation explanation tool is a surrogate rather than a compatible real Activation Oracle. We do not calibrate SAE label quality, logit lens noise, or steering direction quality, so the experiment cannot fully separate interface risk from tool noise. The cleanup step fixes formatting and citation consistency only; it does not rewrite substantive findings. Because reviewers score the cleaned reports, the live burden on an auditor may be higher. Finally, AGORA policy passages support policy audit research but do not by themselves define legal compliance judgments.

\section{Conclusion}

White box evidence packages change how an auditor LLM writes passage anchored policy audit reports, but combined white box evidence alone does not improve grounded audit quality. It increases internal evidence use while weakening grounding and increasing misuse. The evidence is not ignored; it is often read as broad semantic support, even though the stricter patching diagnostic localizes causal support narrowly and with prompt sensitivity. Hybrid packages are more promising in this design, but the result is confounded with package size and should not be read as proof that anchoring alone causes the gain. The shuffled control shows why this distinction matters: plausible reports can still misuse invalid internal evidence, and the report's own cautionary notes do not reliably catch that mismatch. The broader lesson is that internal access is valuable for policy auditing only when it is treated as evidence to be validated, not as authority to be trusted.

\section*{Impact Statement}

This work aims to improve transparency and evidence quality in policy relevant AI auditing. The main risk is over interpreting internal evidence as legal or causal proof. The paper therefore reports validation review, negative controls, and claim boundaries explicitly.

\section*{LLM Usage Statement}

LLMs were used to assist with drafting, editing, formatting, code generation, debugging, summarizing reviewer feedback, and checking clarity. All experimental design decisions, data analysis decisions, validation review, reported numbers, and final claims were reviewed and controlled by the author.

\FloatBarrier

\clearpage

\bibliography{policy_feature_map_refs}
\bibliographystyle{icml2026}

\appendix

\onecolumn

\section{Evidence Condition Details}

The experiment compares evidence interfaces, not retrieval systems. Each condition provides the same policy passage and audit rubric, but changes the auditor facing evidence items. \Cref{tab:appendix-condition-details} expands the condition table used in the main paper.

\begin{table}[H]
    \centering
    \caption{Appendix condition details. Citation rules are part of the interface design: passage spans and evidence entries are separate objects.}
    \label{tab:appendix-condition-details}
    \scriptsize
    \begin{tabular}{@{}p{0.18\textwidth}p{0.26\textwidth}p{0.22\textwidth}p{0.26\textwidth}@{}}
        \toprule
        Condition & Auditor input & Allowed evidence citation & Expected misuse risk \\
        \midrule
        Passage only baseline & Policy passage, source information, and fixed rubric. & Passage spans only. & Claims without passage support may look plausible because no external evidence entries are available. \\
        Black box surface evidence & Passage plus surface observations, issue notes, selected spans, and confidence annotations. & Black box surface entries and passage spans. & Surface notes may encourage generic legal readings, but evidence remains tied to text. \\
        Sparse autoencoder evidence & Sparse feature activations, layers, activated spans, labels, review annotations, and caveats. & Sparse autoencoder entries and passage spans. & Readable feature labels may be treated as policy conclusions. \\
        Logit lens evidence & Intermediate token direction records at selected layers and positions. & Logit lens entries and passage spans. & Token direction evidence may be overread as a causal explanation. \\
        Steering sensitivity evidence & Hidden state direction probes with alpha, token changes, and caveats. & Steering entries and passage spans. & Sensitivity evidence may be mistaken for a deployable control or legal proof. \\
        Activation explanation surrogate & Natural language activation hypotheses derived from local evidence. & Surrogate explanation entries and passage spans. & Natural language hypotheses can sound authoritative despite being unvalidated. \\
        Combined white box evidence & Sparse autoencoder, logit lens, steering, and activation explanation surrogate items. & All combined white box entries and passage spans. & Dense internal evidence can distract from the passage and increase over citation. \\
        Hybrid surface and white box evidence & Black box surface evidence plus combined white box evidence. & Surface entries, white box entries, and passage spans. & Highest average usefulness in the primary review, but still vulnerable to internal evidence overuse. \\
        Raw AutoInterp sparse autoencoder evidence & Raw sparse autoencoder labels and feature references without the surrounding package context. & Raw sparse autoencoder entries and passage spans. & Label readability may hide missing context and caveats. \\
        Shuffled white box relevance control & Combined white box evidence from another case, hidden from the auditor. & Shuffled white box entries and passage spans. & Measures whether plausible internal evidence is trusted when relevance is broken. \\
        \bottomrule
    \end{tabular}
\end{table}

\section{Tool Construction Details}

This appendix defines the evidence artifacts exposed to the auditor. These details matter because the experiment evaluates interfaces, not the intrinsic correctness of any one interpretability tool.

\paragraph{Black box surface evidence.} This interface contains no hidden state information. The passage is split into sentences, and each sentence is scored by an audit cue count $b(s)=\sum_{\tau\in T}\mathbf{1}[\tau \text{ occurs in } s]$. The cue lexicon includes obligation, prohibition, risk, safety, security, transparency, monitoring, assessment, testing, threshold, compliance, enforcement, provider, operator, user, deploy, market, harm, vulnerable, and fundamental rights terms. The five highest scoring sentences become numbered surface evidence entries. Each item stores the sentence index, matched cue count, a supporting passage span, and the caveat that the item is extractive surface evidence.

\paragraph{SAE evidence.} For Gemma 2 2B residual states, Gemma Scope SAE features are summarized as candidate concept activations. For feature $k$ at layer $l$, the package records
\begin{equation}
A_{i,l,k}=\max_{p} a_{i,l,k,p}, \qquad p^{*}_{i,l,k}=\arg\max_{p} a_{i,l,k,p}.
\end{equation}
Features with nonzero activation are sorted by activation score, capped to a compact set, and rendered with layer, feature id, peak token, activated span, model id, SAE release, and caveat. The sparse autoencoder, combined white box, and hybrid interfaces use human revised AutoInterp labels where the review marked the label usable or preserved. The raw AutoInterp sparse autoencoder interface exposes labels without that review layer.

\paragraph{Logit lens evidence.} Logit lens positions are selected from audit cue token offsets in the passage, with a fallback to the final active token. At layers $l\in\{6,12,18,24\}$, the package computes $\ell_{i,l,p}=W_U\mathrm{LN}(h_{i,l,p})$ and records the top five vocabulary tokens and scores. These items are evidence about intermediate token directions, not evidence that the model finally predicts or endorses those tokens.

\paragraph{Steering evidence.} Steering directions are constructed from positive prompts about obligations, risk controls, compliance duties, monitoring, transparency, accountability, enforcement, and harm protection, contrasted with negative prompts describing background text without binding duties. For each layer $l\in\{12,18,24\}$,
\begin{equation}
v_l=\frac{\mu_l^{+}-\mu_l^{-}}{\|\mu_l^{+}-\mu_l^{-}\|_2}, \qquad
h'_{i,l,p}=h_{i,l,p}+0.35\|h_{i,l,p}\|_2v_l.
\end{equation}
The package records base and steered top tokens, whether the top token changes, and the KL divergence between base and steered vocab distributions. It does not run a full generation under intervention.

\paragraph{Activation explanation surrogate.} The activation explanation condition is a conservative surrogate. It converts selected SAE labels, activations, peak tokens, and spans into a short activation meaning hypothesis with explicit caveats. The item is explicitly labeled as a surrogate so that the auditor cannot treat it as a validated Activation Oracle output.

\paragraph{Shuffled white box relevance control.} The control keeps the target policy passage fixed but shifts combined white box packages across cases. Thus passage $x_i$ is paired with a combined white box package from case $j\neq i$ while the auditor sees a normal combined white box shaped package. The control tests whether relevance is being checked or whether plausible internal evidence is being cited because it looks audit like.

\section{Residual Stream Patching Diagnostic}
\label{sec:appendix-patching}

The residual stream patching analysis is a separate behavior locked diagnostic rather than an evidence condition in the report experiment. It uses 60 canonical forced choice microtasks, with 30 authority assignment and 30 deontic force items. Each family is split into 18 train, 6 validation, and 6 test pairs, and each pair is rendered with direct and analyst prompt shells and canonical and swapped answer orders. Both families pass the pre patching behavior gate of rendered accuracy at least 0.70 and at least 20 stable pairs out of 30: authority assignment reaches 0.958 with 29 stable pairs, and deontic force reaches 0.796 with 22. The main layer plot uses the six validation pairs per family under the direct shell and canonical answer order.

For each pair, we replace the corrupted prompt residual vector at a candidate layer and token role with the corresponding clean prompt vector. Gemma 2 2B contributes 26 layers, and the diagnostic tests changed, actor, predicate or modal, threshold or exception, and final prompt token roles. Let $m_{\mathrm{clean}}$, $m_{\mathrm{corrupted}}$, and $m_{\mathrm{patched}}$ denote the correct versus contrast answer letter log probability margins. Patch recovery is
\[
R=\frac{m_{\mathrm{patched}}-m_{\mathrm{corrupted}}}{m_{\mathrm{clean}}-m_{\mathrm{corrupted}}}.
\]
Each pair also contributes 64 random layer and token role trials. A region passes the empirical causal gate only when its mean recovery exceeds the family random region 95th percentile by more than $0.05$ and the lower bound of a 1,000 resample 95\% bootstrap interval exceeds the random region mean.

\Cref{tab:appendix-patching-controls} reports the main diagnostic and rendering controls. In the direct shell, canonical order run, only deontic final prompt token layers 7 and 8 pass; authority assignment has no passing region. Swapping the answer order preserves the deontic signal and adds layer 10, whereas the analyst shell removes every passing region. These controls bound the finding to a rendering sensitive late decision or answer readout signal.

\begin{table}[H]
    \centering
    \caption{Residual stream patching results and rendering controls. Each row reports a selected final prompt token region. The gate threshold is the family random region 95th percentile plus $0.05$, with the additional requirement that the bootstrap lower bound exceed the random region mean.}
    \label{tab:appendix-patching-controls}
    \small
    \begin{tabular}{@{}lllrrrl@{}}
        \toprule
        Family & Rendering & Region & Mean $R$ & 95\% interval & Threshold & Gate \\
        \midrule
        Authority & Direct, canonical & L8 final & 0.0325 & $[0.0106,0.0706]$ & 0.0656 & Fail \\
        Deontic & Direct, canonical & L7 final & 0.1968 & $[0.0302,0.3968]$ & 0.0935 & Pass \\
        Deontic & Direct, canonical & L8 final & 0.1429 & $[0.0220,0.2819]$ & 0.0935 & Pass \\
        Deontic & Direct, swapped & L7 final & 0.2567 & $[0.0996,0.4624]$ & 0.1611 & Pass \\
        Deontic & Direct, swapped & L8 final & 0.2264 & $[0.0886,0.4128]$ & 0.1611 & Pass \\
        Deontic & Direct, swapped & L10 final & 0.1721 & $[0.0331,0.3551]$ & 0.1611 & Pass \\
        Deontic & Analyst, canonical & L12 final & 0.0161 & $[-0.0287,0.0840]$ & 0.0500 & Fail \\
        \bottomrule
    \end{tabular}
\end{table}

\section{Audit Prompt and Report Schema}

The auditor prompt is designed to make traceability a first class output rather than a later reconstruction. The prompt states that the policy passage is the primary source of truth and that evidence items are hypotheses or pointers. The auditor must return a structured report, cite exact short passage spans, avoid invented evidence citations, and avoid causal claims from one tool alone. Each report contains a concise summary; issue findings with actors, obligations or risks, support spans, cited evidence, confidence, and uncertainty; plausible alternative readings; internal evidence that materially influenced the report; cautionary notes about claims that should not be overstated; and an overall confidence score.

The schema separates three evidence objects: the policy passage, condition specific evidence entries, and the auditor's own claims. This separation is what makes shuffled evidence failures visible. A report can have correct passage claims while still citing invalid internal evidence.

\section{Validation Review Protocol}

Validation review has two stages, and five reviewers contribute across gold writing, verification, diagnostic scoring, and agreement checks. First, gold briefs are written from the policy passage and source information without seeing model reports or evidence packages. Each gold brief records issue tags, actors, obligations, support spans, known confounds, and claims to penalize when passage support is missing. Second, diagnostic review scores auditor reports against the gold brief for the four primary conditions: black box surface evidence, combined white box evidence, hybrid surface and white box evidence, and the shuffled white box relevance control. The primary analysis covers all 60 cases. Agreement diagnostics use the original 20 case subset, and optional single tool review scores sparse autoencoder, logit lens, steering sensitivity, activation explanation surrogate, and raw AutoInterp sparse autoencoder evidence on that same subset.

For each reviewed report, the diagnostic vector is
\begin{equation}
h_{i,c}=(h^{\mathrm{corr}}_{i,c},h^{\mathrm{ground}}_{i,c},h^{\mathrm{use}}_{i,c},h^{\mathrm{misuse}}_{i,c}).
\end{equation}
The first three coordinates reward match to the gold brief, while the fourth penalizes invalid evidence use. This means a report can be high quality on passage interpretation and still fail as an audit evidence artifact if $h^{\mathrm{misuse}}_{i,c}$ is high. The shuffled relevance control analysis focuses on exactly this conjunction.

Let $I$ be the reviewed case set and $K$ the four validation review axes. The condition mean for axis $k$ is
\begin{equation}
\bar{h}_{c,k}=\frac{1}{|I|}\sum_{i\in I}h_{i,c,k},
\end{equation}
where $h_{i,c,k}\in\{1,\ldots,5\}$. For pairwise statements relative to the black box surface baseline, we count better, tied, and worse outcomes on the same cases, reversing the interpretation only for misuse because lower misuse is better. We also report paired deltas $\Delta_{i,c,k}=h_{i,c,k}-h_{i,\mathrm{surface},k}$, exact sign tests, and Wilcoxon signed rank normal approximations as descriptive checks. For structural citation uptake, we report $\bar{U}_c=|\dataset|^{-1}\sum_i |Z_{i,c}|$. For the shuffled warning, we count cases satisfying $h_{i,\mathrm{shuffled},\mathrm{correct}}\geq4$ and $h_{i,\mathrm{shuffled},\mathrm{misuse}}\geq4$.

\begin{table}[H]
    \centering
    \caption{Validation review axes and claim boundaries.}
    \label{tab:appendix-human-review-protocol}
    \small
    \begin{tabular}{p{0.19\textwidth}p{0.39\textwidth}p{0.34\textwidth}}
        \toprule
        Axis & Measures & Does not measure \\
        \midrule
        Correctness & Match to the gold issue, actor, and obligation structure. & Whether the target model actually used the cited internal feature causally. \\
        Span grounding & Whether major claims are tied to concrete passage spans. & Whether the cited span is sufficient for a legal decision outside the passage. \\
        Diagnostic usefulness & Whether the report would help an auditor notice the policy issue or gap. & Whether the report is a complete compliance analysis. \\
        Evidence misuse & Whether the report overreads, overcites, or relies on irrelevant internal evidence. & Whether the passage level claim is necessarily wrong. \\
        \bottomrule
    \end{tabular}
\end{table}

\begin{table}[H]
    \centering
    \caption{Agreement diagnostics on the original 20 cases and four primary conditions. QWK denotes quadratic weighted kappa.}
    \label{tab:appendix-iaa}
    \small
    \begin{tabular}{@{}lrrrrr@{}}
        \toprule
        Axis & Reviewer 1 & Reviewer 2 & Exact & Within one & QWK \\
        \midrule
        Correctness & 4.06 & 4.28 & 0.48 & 0.91 & 0.25 \\
        Grounding & 3.25 & 3.71 & 0.48 & 0.91 & 0.50 \\
        Usefulness & 4.08 & 4.00 & 0.53 & 0.93 & 0.46 \\
        Misuse & 2.36 & 2.68 & 0.56 & 0.98 & 0.88 \\
        \bottomrule
    \end{tabular}
\end{table}

\begin{table}[H]
    \centering
    \caption{Optional single tool validation review on the original 20 cases. These results are secondary diagnostics and are not used as the main quality claim.}
    \label{tab:appendix-single-tool}
    \small
    \begin{tabular}{@{}p{0.34\textwidth}rrrr@{}}
        \toprule
        Condition & Correct & Ground & Useful & Misuse \\
        \midrule
        Sparse autoencoder evidence & 3.45 & 3.15 & 3.00 & 3.25 \\
        Logit lens evidence & 2.40 & 1.25 & 1.95 & 4.00 \\
        Steering sensitivity evidence & 3.35 & 3.25 & 3.00 & 3.00 \\
        Activation explanation surrogate & 3.30 & 3.00 & 3.00 & 2.70 \\
        Raw AutoInterp sparse autoencoder evidence & 3.30 & 1.25 & 2.00 & 4.00 \\
        \bottomrule
    \end{tabular}
\end{table}

\section{All Condition Structural Summary}

\begin{figure}[H]
    \centering
    \includegraphics[width=0.94\textwidth]{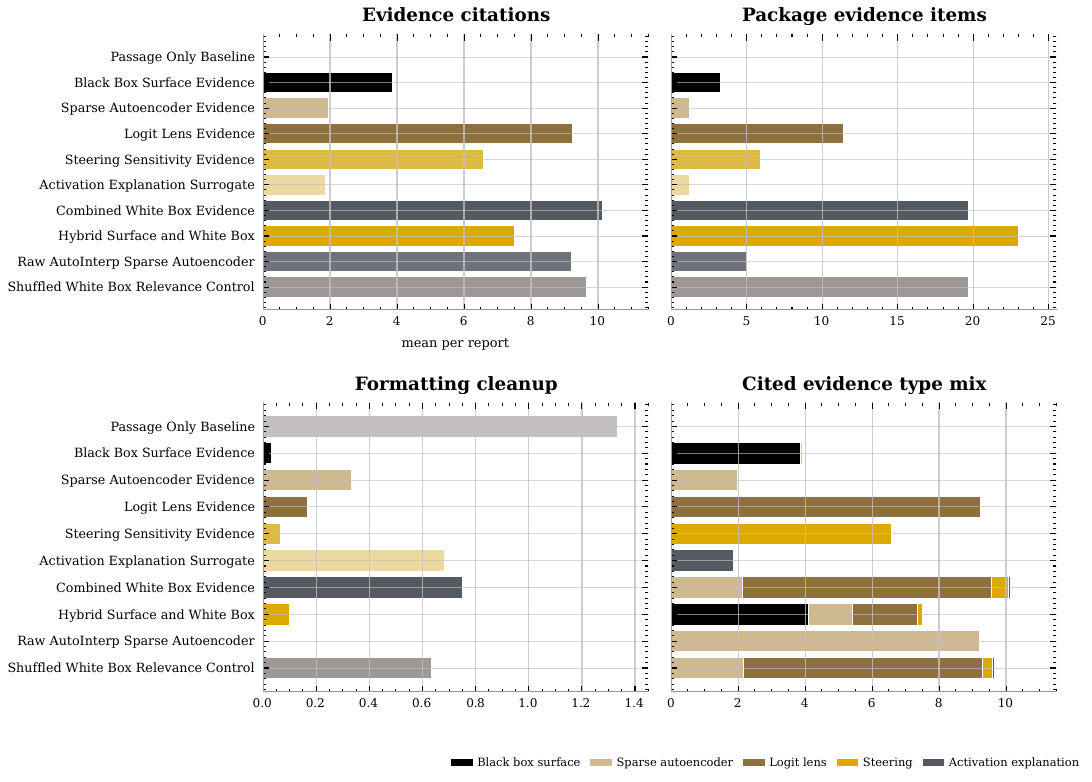}
    \caption{Structural metrics across all ten evidence conditions. Single tool conditions change citation behavior and formatting burden, but these measures are not validation scores. Logit lens evidence and raw AutoInterp sparse autoencoder evidence show high citation volume, while the shuffled white box relevance control shows that a combined white box shaped package can be cited heavily even when relevance is broken.}
    \label{fig:appendix-structural-summary}
\end{figure}

\section{Case Study 1: Hybrid Anchoring Works}

This case uses Articles 29 and 30 of the Internet Information Service Algorithmic Recommendation Management Provisions. The passage contains two distinct policy issues: confidentiality duties for bodies and personnel involved in supervision, and a complaint or report mechanism for organizations and individuals.

\begin{quote}
\small
Article 29: Related bodies and personnel participating in algorithmic recommendation service security assessment, supervision, and inspection shall maintain confidentiality of the personal private information, personal information, and commercial secrets they learn when exercising their duties and responsibilities, they may not disclose, sell, or illegally provide it to other persons. Article 30: Where any organization or individual discovers acts violating these Provisions, they may file a complaint or report with cybersecurity and informatization departments and relevant departments. Departments receiving complaints or reports shall handle them timely and according to the law.
\end{quote}

The gold brief identifies confidentiality in supervision, complaint and report rights, and complaint handling obligations. \Cref{tab:appendix-cn21-case} shows why the hybrid surface and white box interface is better for this case. The important point is not that white box evidence alone solves the interpretation. Combined white box evidence narrows the report to confidentiality and legal liability, while the hybrid interface keeps the surface anchors that identify both Article 29 and Article 30.

\begin{table}[H]
    \centering
    \caption{Case study 1 diagnostic comparison for a Chinese algorithmic recommendation provision on confidentiality and complaint handling. Scores are correctness, grounding, usefulness, and misuse. Lower misuse is better.}
    \label{tab:appendix-cn21-case}
    \scriptsize
    \setlength{\tabcolsep}{3pt}
    \begin{tabular}{@{}p{0.17\textwidth}p{0.27\textwidth}p{0.21\textwidth}p{0.23\textwidth}@{}}
        \toprule
        Condition & Report behavior & Cited evidence pattern & Validation interpretation \\
        \midrule
        Black box surface evidence & Identifies confidentiality and complaint handling, but grounding is not fully comprehensive. & Uses selected surface spans from the passage. & 5, 4, 5, 1. Strong passage anchored baseline. \\
        Combined white box evidence & Captures confidentiality but weakens grounding and usefulness relative to the surface baseline. & Uses logit lens and steering evidence around selected token positions. & 5, 3, 4, 2. Related internal evidence is not enough to preserve full passage structure. \\
        Hybrid surface and white box evidence & Covers confidentiality, nondisclosure, complaint or report rights, and handling by departments. & Uses black box surface anchors while white box evidence remains secondary. & 5, 5, 5, 2. Highest scored result in this case; passage anchors preserve both Articles 29 and 30. \\
        Shuffled white box relevance control & Produces a partially correct confidentiality finding but cites shuffled transparency evidence. & Uses shuffled SAE, logit lens, and steering evidence. & 4, 2, 3, 5. The internal evidence is mismatched even though the passage claim is partly correct. \\
        \bottomrule
    \end{tabular}
\end{table}

\section{Case Study 2: White Box Can Misframe Legal Category}

This case uses an EU AI Act provision on high risk classification exceptions. The passage is legally delicate because it is not a generic operator duty. It specifies when an Annex III system is not considered high risk, while preserving an override for profiling of natural persons.

\begin{quote}
\small
By derogation from paragraph 2, an AI system referred to in Annex III shall not be considered to be high risk where it does not pose a significant risk of harm to the health, safety or fundamental rights of natural persons, including by not materially influencing the outcome of decision making. The first subparagraph shall apply where any listed condition is fulfilled, including a narrow procedural task, improving a previously completed human activity, detecting decision making patterns without replacing or influencing the prior human assessment without proper human review, or performing a preparatory task. Notwithstanding the first subparagraph, an AI system referred to in Annex III shall always be considered to be high risk where the AI system performs profiling of natural persons.
\end{quote}

The gold brief identifies a classification exception, a significant risk threshold, decision making influence, Annex III scope, and a profiling override. \Cref{tab:appendix-eu21-case} shows a limitation of the hybrid result: even the hybrid surface and white box interface can shift to imprecise legal wording. This case keeps the paper's claim boundary intact. Hybrid evidence is promising on average, not a guarantee of legal category precision.

\begin{table}[H]
    \centering
    \caption{Case study 2 diagnostic comparison for an EU AI Act provision on high risk classification exceptions. Scores are correctness, grounding, usefulness, and misuse. Lower misuse is better.}
    \label{tab:appendix-eu21-case}
    \scriptsize
    \setlength{\tabcolsep}{3pt}
    \begin{tabular}{@{}p{0.17\textwidth}p{0.27\textwidth}p{0.21\textwidth}p{0.23\textwidth}@{}}
        \toprule
        Condition & Report behavior & Cited evidence pattern & Validation interpretation \\
        \midrule
        Black box surface evidence & Captures non high risk exceptions and the profiling carve out. & Uses surface evidence tied to passage spans. & 4, 4, 4, 1. Correct but could state more clearly that this is classification criteria. \\
        Combined white box evidence & Preserves the broad classification topic but can still blur the legal category. & Uses logit lens evidence around selected token positions. & 4, 3, 4, 2. Partially correct, but internal evidence is weak support for the legal distinction. \\
        Hybrid surface and white box evidence & Captures conditions and profiling but uses imprecise low risk wording. & Mainly uses surface passage evidence despite receiving white box evidence. & 4, 4, 4, 2. Hybrid helps but the legal category is less precise than not high risk. \\
        Shuffled white box relevance control & Infers a risk assessment gap and misses part of the mandatory profiling carve out. & Uses shuffled deployment evidence and unrelated logit lens evidence. & 3, 2, 2, 5. Control evidence is mismatched and the key exception structure is weakened. \\
        \bottomrule
    \end{tabular}
\end{table}

\section{Shuffled Control Failure Table}

\Cref{tab:appendix-c9-failures} lists representative shuffled white box relevance control reports from the 39 cases with correctness at least 4 and evidence misuse at least 4. These are the most important governance failures because a reviewer who checks only passage level correctness could miss the invalid internal evidence use.

\begin{table}[H]
    \centering
    \caption{Representative shuffled white box relevance control cases with plausible correctness and high evidence misuse.}
    \label{tab:appendix-c9-failures}
    \scriptsize
    \begin{tabular}{@{}p{0.07\textwidth}p{0.14\textwidth}p{0.24\textwidth}p{0.15\textwidth}p{0.25\textwidth}@{}}
        \toprule
        Case & Source & Gold issue & Shuffled evidence cited & Reviewer finding \\
        \midrule
        CN 12 & Chinese AR Provisions & User notice, transparency, service mechanism disclosure. & Logit lens. & Correctness 4 and misuse 5: passage overlap is plausible, but internal evidence is invalid support. \\
        CN 13 & Chinese AR Provisions & Non targeted option, switch off right, tag controls, explanation for major rights impacts. & SAE and logit lens. & Correctness 4 and misuse 5: readable shuffled evidence makes the report look more grounded than it is. \\
        CN 14 & Chinese AR Provisions & Minor protection, beneficial content, harmful content restriction, addiction prevention. & SAE and logit lens. & Correctness 4 and misuse 5: surface conclusion is close, while cited internal evidence comes from another case. \\
        CN 21 & Chinese AR Provisions & Confidentiality, complaint or report right, complaint handling obligation. & SAE, logit lens, and steering. & Correctness 4 and misuse 5: the report recovers part of the passage but relies on shuffled transparency evidence. \\
        CN 24 & Chinese AR Provisions & Filing fraud sanctions, filing cancellation, administrative penalties. & SAE and logit lens. & Correctness 4 and misuse 5: close to the passage, but control evidence should not be treated as support. \\
        CN 7 & Chinese AR Provisions & Information security, unlawful content handling, records preservation. & SAE. & Correctness 4 and misuse 5: generic SAE labels appear persuasive despite being mismatched. \\
        EU 182 & EU AI Act & Right to explanation and high risk AI deployer obligation. & SAE and logit lens. & Correctness 4 and misuse 5: primary obligation is plausible, but control evidence is mismatched. \\
        EU 231 & EU AI Act & Technical documentation, validation, logs, cybersecurity measures. & Logit lens. & Correctness 4 and misuse 5: mostly correct passage summary with invalid internal citations. \\
        EU 39 & EU AI Act & Human oversight, role allocation, and monitoring obligations. & Logit lens and steering. & Correctness 4 and misuse 5: the report sounds grounded while internal evidence relevance is broken. \\
        GOV 1396-4 & Governance principles & Governance process, accountability, or safety framework requirements. & SAE and logit lens. & Correctness 4 and misuse 5: a broad governance theme survives, but the cited evidence package is from another case. \\
        \bottomrule
    \end{tabular}
\end{table}

\section{Validation Review Score Distributions}

\Cref{tab:appendix-score-distributions} reports the score distributions behind the 60 case means. Each vector lists counts for scores 1 through 5. The distributions make the same claim boundary visible: the hybrid interface has many high usefulness scores, but it also shifts misuse upward; the shuffled relevance control remains often plausible on correctness while all shuffled reports score 5 on misuse.

\begin{table}[H]
    \centering
    \caption{Expanded validation review score distributions over 60 cases. Each vector is counts for scores 1/2/3/4/5.}
    \label{tab:appendix-score-distributions}
    \scriptsize
    \setlength{\tabcolsep}{3pt}
    \begin{tabular}{@{}p{0.28\textwidth}rrrr@{}}
        \toprule
        Condition & Correct & Ground & Useful & Misuse \\
        \midrule
        Black box surface evidence & 0/0/0/19/41 & 0/0/0/29/31 & 0/0/0/19/41 & 60/0/0/0/0 \\
        Combined white box evidence & 0/0/0/24/36 & 0/4/37/19/0 & 0/0/0/60/0 & 0/30/30/0/0 \\
        Hybrid surface and white box evidence & 0/0/0/18/42 & 0/0/4/22/34 & 0/0/0/5/55 & 0/52/8/0/0 \\
        Shuffled white box relevance control & 0/1/20/39/0 & 0/27/26/7/0 & 1/20/39/0/0 & 0/0/0/0/60 \\
        \bottomrule
    \end{tabular}
\end{table}

\section{Reproducibility Notes}

The artifact repository at \url{https://github.com/GRAIL-center/agora-mi} contains the 600 JSONL reports and the structural, validation, paired, agreement, and single tool summaries used in the paper. The appendix case studies use the named gold briefs, review sheets, and final reports.

Citation validity is a structural check:
\begin{equation}
V_{i,c}=\mathbf{1}\left[ Z_{i,c}\subseteq \{id(e):e\in\package_{i,c}\}\right].
\end{equation}
The value 1.0 means that cited identifiers occur in the supplied package after parsing, schema, confidence, and citation cleanup. It does not establish substantive relevance, so the shuffled relevance control can have perfect citation validity and high human scored misuse. Cleanup does not alter the passage, evidence package, or report findings.

\end{document}